\documentclass[%
 reprint,
 amsmath,amssymb,
 aps,
longbibliography
]{revtex4-2}

\usepackage{graphicx}
\usepackage{dcolumn}
\usepackage{bm}
\usepackage{breakurl}
\usepackage[dvipsnames]{xcolor}
\usepackage{float}
\usepackage{ulem}
\usepackage{placeins}

\begin{document}

\preprint{APS/123-QED}

\title{Quantum amplitude damping for solving homogeneous linear differential equations: A noninterferometric algorithm}

\author{Jo\~ao H. Romeiro$^1$}\email{joao.henrique.alves@alumni.usp.br}
\author{Frederico Brito$^2$}\email{fbb@ifsc.usp.br}
\affiliation{$^1$Department of Electrical and Computer Engineering, S\~ao Carlos School of Engineering, University of S\~ao Paulo, S\~ao Carlos, Brazil.\\ ~$^2$Instituto de F\'isica de S\~ao Carlos, Universidade de S\~ao Paulo, CP 369, 13560-970, S\~ao Carlos, S\~ao Paulo, Brazil.}


\begin{abstract}

In contexts where relevant problems can easily attain configuration spaces of enormous sizes, solving Linear Differential Equations (LDEs) can become a hard achievement for classical computers; on the other hand, the rise of quantum hardware can conceptually enable such high-dimensional problems to be solved with a foreseeable number of qubits, whilst also yielding quantum advantage in terms of time complexity. Nevertheless, in order to bridge towards experimental realisations with several qubits and harvest such potential in a short-term basis, one must dispose of efficient quantum algorithms that are compatible with near-term projections of state-of-the-art hardware, in terms of both techniques and limitations. As the conception of such algorithms is no trivial task, insights on new heuristics are welcomed. This work proposes a novel approach by using the Quantum Amplitude Damping operation as a resource, in order to construct an efficient quantum algorithm for solving homogeneous LDEs. As the intended implementation involves performing Amplitude Damping exclusively via a simple equivalent quantum circuit, our algorithm shall be given by a gate-level quantum circuit -- predominantly composed of elementary 2-qubit gates -- and is particularly nonrestrictive in terms of connectivity within and between some of its main quantum registers. We show that such an open quantum system-inspired circuitry allows for constructing the real exponential terms in the solution in a non-interferometric way; we also provide a guideline for guaranteeing a lower bound on the probability of success for each realisation, by exploring the decay properties of the underlying quantum operation.
\end{abstract}

\maketitle

\section{Introduction}

The foreseeable reality of quantum technologies operating in a fully coherent manner imposes the challenge of being able to explore the underlying quantum phenomena as resources, in order to enhance processes of interest effectively. Results ranging from quantum information processing, with the famous Shor's quantum factoring algorithm \cite{365700}, to quantum heat engines \cite{PhysRevLett.122.110601} help foster those expectations. Nevertheless, identifying the quantum protocol that can conceptually surpass its classical counterparts -- whilst also being adapted to near-term hardware in terms of employable paradigms and experimentally achievable techniques -- has proven to be a generally non-trivial task, and the development of new strategies is welcomed.

Solving linear differential equations is one of the important examples for which classical computers can struggle to perform the task, particularly in contexts where one can quite easily reach very large configuration spaces, as the size of the problem increases, and a quantum computer can offer a real advantage. Indeed, several efficient quantum algorithms have already been proposed, for nonlinear differential equations \cite{leyton2008quantum, lloyd2020quantum, liu2021efficient} and linear differential equations \cite{xin2020quantum}. Whereas the former set of examples allow for solving wide-ranging classes of equations, they lack a straightforward quantum circuit form, which could make its translation to an experimental implementation especially difficult. The latter example, on the other hand, has been experimentally verified for a few qubits and is given by an equivalent quantum circuit, although it is still described in a high-level manner with abstract, potentially many-qubits quantum gates, which could hinder scaling to larger experimental setups with connectivity and other hardware limitations.

Here, we introduce an efficient quantum algorithm for solving Homogeneous Linear Differential Equations (HLDEs), based on a novel strategy to construct the correct answer through the heuristic use of Amplitude Damping (via an equivalent quantum circuit). Indeed, instead of using interference to suppress the wrong outcomes, we explore the decay properties of said quantum operation to favour the correct answer. This approach allows for an equivalent quantum circuit which -- to the exception of a Hamiltonian simulation sub-module -- is given at a decomposed gate-level description, hence being significantly more friendly towards a potential experimental implementation, as well as towards application to use-cases of interest in foreseeable scalable hardware. The algorithm will solely use a Quantum Phase Estimation (QPE) module followed by a section exclusively composed of elementary 2-qubits gates (CNOTs and parameterized controlled 1-qubit rotations) with an extra auxiliary register, responsible for implementing the non-unitary aspect of the calculations via a subsequent set of measurements. Also notably, this chain of controlled gates only requires entanglement to be created within pairs of qubits (one for each register), which can be less restrictive in terms of qubit connectivity on a quantum chip and facilitate the underlying mapping, as well as scalability. 

The use of quantum operations as direct resources to a specific protocol has already been explored in \cite{mahadev2018classical}, for classical verification of quantum computations. Here, we exemplify their potential as building blocks to quantum algorithms as well, specifically as a versatile strategy for realising non-unitary manipulations. Indeed, this could potentially be only an example from a family of circuit-defined algorithms inspired by open quantum system dynamics.

\section{Problem delimitation}

Our goal will be to solve Homogeneous Linear Differential Equations (HLDE), i.e. for a given set of initial conditions $x_0 \in \mathbb{C}^N$, time value  $t\in \mathbb{R}_+$ and $A \in M_N(\mathbb{C})$, we would like to calculate an unknown vector $x(t) \in \mathbb{C}^N$ that obeys the following:
\begin{equation}
\label{eq:problemi}
\begin{cases}
\frac{dx(t)}{dt} = Ax(t) \\
x(0) = x_0
\end{cases}
\end{equation}

In addition, a couple of extra conditions are imposed on matrix $A$. The primary extra condition is that $A$ must be Hermitian, specifically because the existence of a basis of eigenvectors of $A$ on $\mathbb{C}^N$ - yielded by the Spectral Theorem - is a chore aspect on which our framework stands, as in \cite{harrow2009quantum}. The generalisation to non-Hermitian cases can be achieved by iterating the presented algorithm and a Hamiltonian Simulation module for the anti-Hermitian component, via the Lie-Trotter formula \cite{berry2014high}. 

As a second restriction, we will initially assume that all eigenvalues of $A$ are strictly positive, as we deem this particular case to be illustrative of how our algorithm operates. This constraint will be promptly released in Appendix \ref{APB}, where it is shown that the more general case can be achieved by the same quantum circuit layout and size by simply redefining a few gate parameters.

Finally, we will hereby let the dimension $N$ be a power of $2$ for simplicity while describing the algorithm, given that the generalisation to other cases should be trivial.

\begin{figure*}[hbt!]
    \centering
    \includegraphics[scale = 0.3]{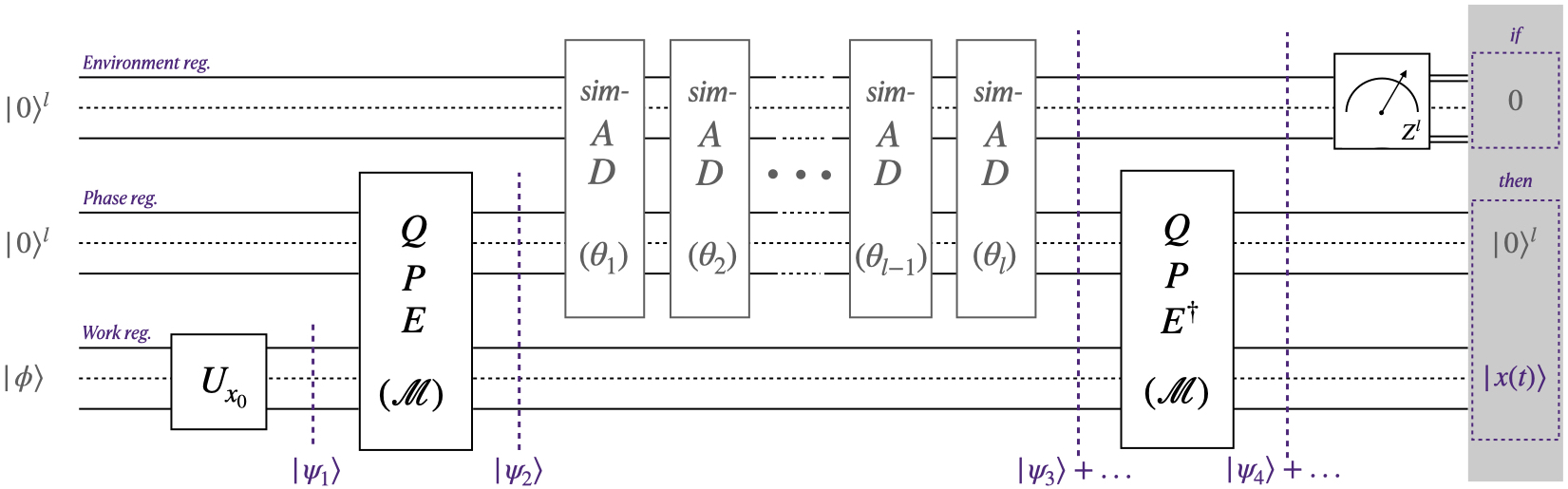}
    \caption{Proposed quantum circuit for solving HDLEs when $A$ is Hermitian. It includes three main quantum registers: the work register (of size $n\equiv \log_2(N)$), the phase register and the environment register (both of size $l\equiv \log_2(L)$, stipulated by the intended precision); the circuit is divided into 4 modules plus measurements at the end.}
        \label{fig:Fig1}
\end{figure*}

\section{Algorithm's overview}
  
\subsection{Algorithm's framework}

The HDLE problem from  Eq. \ref{eq:problemi} has an analytical solution given by $x(t) = e^{At}x_0$. One way of going about solving this equation would be to calculate the matrix exponential $e^{At}$, for instance by finding the decomposition $A = VDV^{-1}$, where $D$ is diagonal, and calculating $e^{At} = Ve^{Dt}V^{-1}$, which can be achieved efficiently if one is able to calculate the exponential of a scalar efficiently. Nevertheless, any method for a classical computer to approximate $e^{At}$ in general requires time that scales at least as $O(N^3)$ \cite{dubious}, which dominates the total complexity since the subsequent matrix-vector multiplication $e^{At}x_0$ only takes quadratic time.

Alternatively, one could rewrite the right-hand side of the expression as $x(t) = \sum_{i=1}^Ne^{a_it}\tilde{c}_i v_i $, where $\{v_i\}_{1\leq i\leq N}$ and $\{a_i\}_{1\leq i\leq N}$ are the normalized eigenvectors and associated eigenvalues of $A$, respectively, and $\tilde{c}_i\equiv  \langle x_0 , v_i\rangle_{\mathbb{C}^N}$; this decomposed analytical expression provides guideline for our quantum algorithm. In what follows, we will consider that $(v_i)_{1\leq i\leq N}$ and $(a_i)_{1\leq i\leq N}$ are labelled in such a way that the eigenvectors are sorted in ascending order, i.e. $a_N$ is an instance of the largest eigenvalue and thus $||A||_2 = a_N$. For now on, whenever we evoke $||A||$, it will correspond to the 2-norm, unless an alternative sub-index is given.

For the sake of translating the basics of our problem into quantum, we now define the following normalized state vectors, which are respectively proportional to the initial conditions and the solution:

\begin{equation}
\begin{cases}
\label{eq:stdef}
|x_0\rangle \equiv \frac{1}{||x_0||}\sum_{i=1}^N(x_0)_i|i\rangle,\\
|x(t)\rangle \equiv \frac{1}{\sqrt{\gamma(t)}} \sum_{i=1}^{N}e^{-(a_N-a_i)t}c_i|v_i\rangle.

\end{cases}
\end{equation}

For the $|x(t)\rangle$ expression, $\gamma(t)$ is a normalisation factor  (with $\gamma(0)=1$ and $\gamma(t)\leq1$), $\forall i\in \{1,2...N\} \quad c_i \equiv \tilde{c}_i/||x_0||$ and $ |v_i\rangle = \sum_{j=1}^N(v_i)_j|j\rangle$. Note that the expression for $|x(t)\rangle$ is chosen such that it includes a global factor $e^{-a_Nt}$, in such a way that the explicit exponential terms in Eq. \ref{eq:stdef} are either vanishing or constant in $t$. This choice is motivated by the gamut of operations that are more intuitively implemented via Amplitude Damping -- the algorithm's main resource -- due to its generally dissipative nature. 

Our global goal will be to initialise a quantum register composed of $n\equiv \log_2(N)$ qubits (the work register) with the state $|x_0\rangle$ and to apply a series of operations in order to approximate $|x(t)\rangle$. We hereby identify the need to perform a non-unitary operation locally to the associated quantum register, given the scaling by real exponential terms that must be performed. For equivalent reasons, we will allow for a potentially non-unit {\bf probability of success $p_S \in ]0,1]$}, i.e. the algorithm might fail for a finite set of repetitions, as long as it also outputs classical information telling if the algorithm succeeds, which will come in the form of a measure to a set of auxiliary qubits at the end of the calculations.

We also explore the following constraint: for some given $A$ and $x_0$, one should be able to affirm that the algorithm succeeds for any $t$ with at least some non-zero probability, i.e. we should be able to put a $t$-independent positive lower bound on $p_S$. Generally, for all the instances that do not inherently meet the aforementioned criterion, we propose a very simple alternative method on Section \ref{APC}, that can also be used in order to boost the probability of success $p_S$ even in cases where it already holds a positive lower-bound. One could also aim at adapting our algorithm to an Amplitude Amplification \cite{brassard2002quantum} framework in order to optimise the number of required repetitions.

Finally, if the algorithm succeeds, then some global characteristic $F(x)= \langle x(t)|M|x(t)\rangle$ can be extracted through the application of the corresponding observable $M$. Alternatively, the whole state can be approximated through several repetitions and Quantum Tomography \cite{altepeter2005photonic, torlai2018neural}. Under any of these circumstances, all findings are linked to the original problem through the explicit relation $x(t) \dot{=} \sqrt{\gamma(t)}\:\big|\big|x_0\big|\big|e^{||A||t}\:|x(t)\rangle$. 

\subsection{Equivalent quantum circuit}

The overall intuition behind the algorithm's functioning can be understood as follows: after preparing the work register in the initial condition state $|x(0)\rangle$, we let it non-unitarily evolve to $|x(t)\rangle$ by creating a set of decay effects to the qubits on an auxiliary register. The probability of such decays are conditioned on the eigenvalues $(a_i)_{1\leq i \leq N}$ and on $t$, and can thus be used to create the $(e^{-(a_N-a_i)t})_{1\leq i \leq N}$ terms. Measuring the auxiliary register and verifying that no decay occurred from the other registers (i.e. measurement yields $|0\rangle$) will then project the state of the work register into the desired (potentially approximated) $|x(t)\rangle$ state.

The resulting implementation is given by the quantum circuit in Fig. \ref{fig:Fig1} and is composed of four main modules, which will be covered by order of appearance in the following sub-sections.

\subsubsection{IC preparation}

As previously stated, the first stage of our algorithm corresponds to loading the initial conditions (IC) vector into the work register, resulting in $|\psi_1\rangle \equiv |x_0\rangle$ as given by Eq. \ref{eq:stdef}. This is a reoccurring starting module for several algorithms aimed at solving similar problems \cite{harrow2009quantum, xin2020quantum} and also in Quantum Machine Learning applications \cite{biamonte2017quantum}.

\subsubsection{Quantum phase estimation}
\label{sec:QPE}

In order to create the dependency of the upcoming decay effects on the eigenvalues of $A$, representations of $(a_i)_{1\leq i \leq N}$ must be loaded into an auxiliary register. First, we define the operators $\bar{A} \equiv A/||A||$ and the unitary operator $\mathcal{M}\equiv e^{-i2\pi\bar{A}}$. The second stage of our algorithm is the Quantum Phase Estimation (QPE) \cite{cleve1998quantum} of $\mathcal{M}$, i.e. the phase estimation module is constructed via controlled-$\mathcal{M}$ operations. The register used to code the phases (hereby called phase register) is of size $l \in \mathbb{N}^*$, thus individually corresponding to a Hilbert space of size $L\equiv 2^l$. The output of the quantum circuitry presented so far is:
\begin{equation}
\label{eq:QPEform1}
|\psi_2\rangle \equiv  \sum_{i=1}^{N}c_i|\tilde{\phi}_i\rangle|v_i\rangle, 
\end{equation}
where the state vectors $(|\tilde{\phi}_i\rangle)_{1\leq i\leq N}$ code the phases $(e^{i2\pi \phi_i})_{1\leq i\leq N} \equiv (e^{-i2\pi a_i/||A||})_{1\leq i\leq N}$, the spectrum of $\mathcal{M}$, in a potentially approximated manner, and are explicitly given by \cite{nielsen2002quantum}:

\begin{equation*}
    |\tilde{\phi}_i\rangle = \frac{1}{L}\sum_{k,j=0}^{L-1}e^{-\frac{i2\pi kj}{L}} e^{i2\pi k \phi_i}|j\rangle
\end{equation*}

with $\phi_i \equiv 1 - a_i/||A||$. It should be noted that at least $\tilde{\phi}_N$ is an exact approximation, that is, $|\tilde{\phi}_N\rangle = |0\rangle^l$.

\begin{figure}[!hbt]
    \centering
    \includegraphics[scale = 0.26]{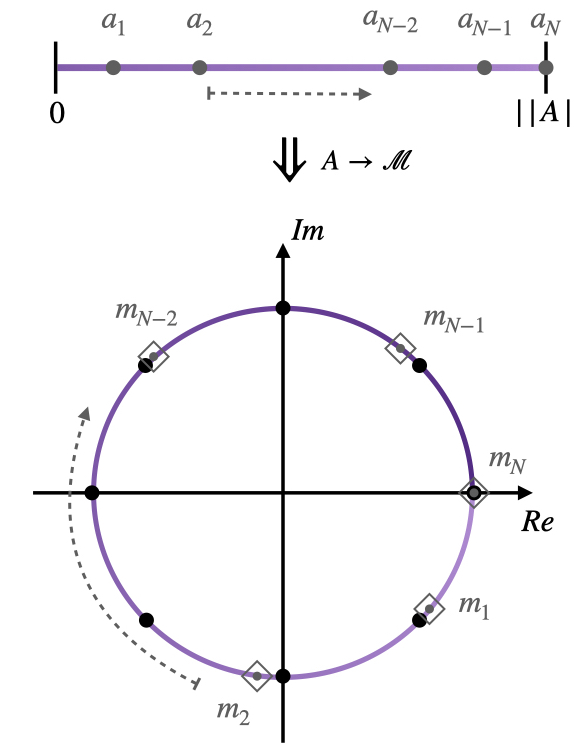}
    \caption{Illustrative example of the eigenvalues mapping between $A$ and $\mathcal{M}$, given by $(m_i)_{1\leq i\leq N} \equiv (e^{-i2\pi a_i/||A||})_{1\leq i\leq N}$. The eigenvalues of $A$ are wrapped in a clockwise manner around the unit circle, as indicated by the dashed arrow. The black filled dots around the circle represent all phases that can be exactly represented by a single computational state in the phase register, in this example for $L=8$. These can also be regarded as the allowed levels of damping that can later be performed by the \textit{sim}-AD chain. The largest eigenvalue of $A$ is always set to $m_N = 0$. For all other eigenvalues, a mismatch with relation to the black dots will engender a superposition of computational states in the phase register, which is a mechanism of error in the algorithm.} 
    \label{fig:Fig3_ant}
\end{figure}

Indeed, it can be verified that by rewriting $|\tilde{\phi}_i\rangle$ as a superposition $\sum_{j=0}^{L-1}\beta_{i;j}|j\rangle$, then $|\beta_{i;j}|$ takes a high value {\it  iff} $j \approx L(1-a_i/||A||) \pmod L$. If this approximation is valid for all $i\in \{1, 2...N-1\}$ and for some $j\in \{0, L-1\}$ each, then the right-hand side of Eq. \ref{eq:QPEform1} can be simplified as:
\begin{equation}
\label{eq:QPEform2}
|\psi_2\rangle \approx
\sum_{i=1}^{N}c_i|  d_i \rangle |v_i\rangle
\end{equation}

Where $d_i \equiv \lfloor L(1-a_i/||A||)\rceil \pmod L$, with the $\lfloor.\rceil$ notation indicating the closest integer value. The validity of this approximation depends on the distribution of eigenvalues of $\mathcal{M}$ around the unitary circle, but is also a fair assumption if the phase register is sufficiently large, i.e. allowing for more possible values of $j$. This is intuitively depicted in Figure \ref{fig:Fig3_ant}.

We shall continue by using Eq. \ref{eq:QPEform2} as the expression for $|\psi_2\rangle$, given that much more substantial analysis over the associated error will be provided in Appendix \ref{tech_err}.

\subsubsection{\textit{sim}-AD modules}

The subsequent section in our quantum circuit concentrates the heuristic use of the Quantum Operation known as Amplitude Damping (AD). This operation is an important tool for describing processes that dissipate energy in open quantum systems, where a (ground) state amplitude is favoured against others due to their decay. Here we explore and combine several of these amplitude decay effects in order to construct the exponential terms $(e^{-(a_N-a_i)t})_{1\leq i\leq N}$ that appear on the solution. As shown on Section \ref{APC}, such a feature can also be used to single out the correct answer through the amplitude decay of the wrong outcomes, differing from the standard approaches which rely on interferometric schemes.

\begin{figure}
    \centering
    \includegraphics[scale = 0.3]{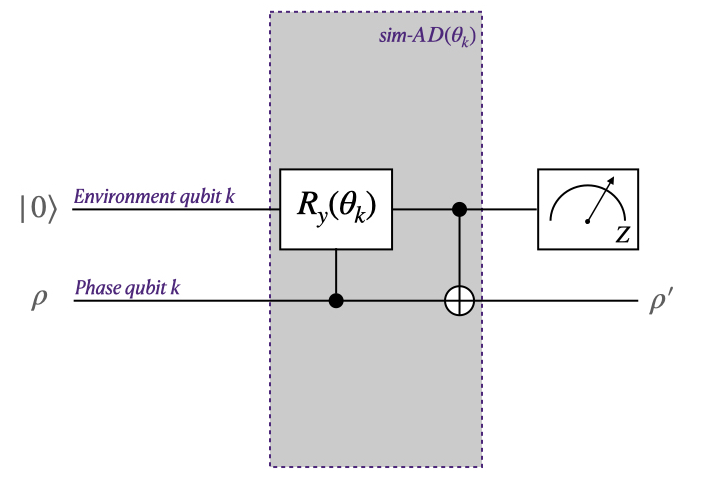}
    \caption{White-box representation of a \textit{sim}-AD module. This simple quantum circuitry realises the Amplitude Damping quantum operation on $\rho$, whereas the auxiliary upper qubit plays the role of an environment. $R_y(\theta_k)\equiv e^{-i\theta_k\sigma_y/2}$, where $\sigma_y$ is the complex Pauli Matrix.}
    \label{fig:Fig2}
\end{figure}

Before proceeding, it should be made clear that we do not demand the manipulation of an actual open quantum system in order to implement our algorithm -- such a requirement would not be desirable in several physical implementations since it could be hard ensuring that only such a process would be present. Nevertheless, we deem the Amplitude Damping operation to be heuristic in our task, as we shall use the quantum circuit shown in Fig \ref{fig:Fig2}, which is capable of \textit{simulating} this effect \cite{nielsen2002quantum}, particularly if it were happening under completely controlled circumstances: hence our usage of the term \textit{sim}-AD.

If the measurement result from Fig. \ref{fig:Fig2} is unknown, then the output for an input density operator $\rho$ follows the operator-sum expression given by $\rho' = E^k_0\rho {E_0^k}^\dagger{} + E_1^k\rho {E_1^k}^\dagger{}$, where: 
\begin{equation}
    \label{kraus}
    \begin{cases}
    
    E^k_0 = 
    \begin{bmatrix}
    1 & 0 \\
    0 & \cos(\theta_k/2)
    \end{bmatrix} \\
    \\
    
     E^k_1 = 
    \begin{bmatrix}
    0 & \sin(\theta_k/2)\\
    0 & 0
    \end{bmatrix}

    \end{cases}
\end{equation}

Correspond to the Amplitude Damping's Kraus operators \cite{wilde2011classical}. However, if the measurement result is known and equal to $0$, then the final state is proportional to $E^k_0\rho {E^k_0}^{\dagger{}}$, i.e. if the input is a pure state $|\psi \rangle$, the output is $\frac{E^k_0|\psi\rangle}{\sqrt{\langle \psi |{E^k_0}^\dagger{}E^k_0| \psi \rangle}}$.

We now introduce a third and last register, called environment register, which should also be composed of $l$ qubits. Each $AD_k$ module from Fig. \ref{fig:Fig1} corresponds to the 2-gates circuit from Fig. \ref{fig:Fig2} applied to the $k$-th phase qubit and the $k$-th environment qubit, while the state of all the other qubits is trivially operated upon. The parameters $(\theta_k)_{0 \leq k \leq l-1}$ are selected such that $\cos(\theta_k/2) \equiv e^{-||A||2^kt/L}$, leading to the following set of design equations:
\begin{equation}
    \label{eq:thetas}
    \begin{cases}
    \forall k\in \{0,1...l-1\} \: \theta_k \equiv \arccos(2e^{-||A||2^{k+1}t/L}-1) 
    \end{cases}
\end{equation}

Furthermore, if the measurements for each \textit{sim}-$AD_k$ module outputs $0$, the resulting (non-unitary) operator applied to the phase register is $G_{0} \equiv \bigotimes_{k=0}^{l-1}E_0^k$. An expression for this operator is explicitly calculated in Appendix \ref{closerlook}, where it is shown to be equal to $ \sum_{j=0}^{L-1}e^{-||A||tj/L}|j\rangle \langle j|$. 

Finally, we can look into the outcome of applying the chain of \textit{sim}-$AD$ modules to $|\psi_2\rangle$, if the measurement outcomes are all $0$:
\begin{equation}
\label{eq:expAPA}
|\psi_3\rangle \approx \frac{1}{\sqrt{\gamma(t)}}  
\sum_{i=1}^{N}c_ie^{-(a_N-a_i)t}  | d_i \rangle|v_i\rangle
\end{equation}

Where, once again, a more thorough step-by-step calculation is given in Appendix \ref{closerlook}. The explicit value of the normalisation constant is $\gamma(t) \equiv \sum_{i=1}^{N}|c_i|^2e^{-2(a_N-a_i)t}$.

Intuitively, if each local $|1\rangle$ state is interpreted as an excitation \footnote{note that this is merely an interpretation and does not rely on the physical realisation of $|1\rangle$ being an excited state, as dictated by the underlying hardware}, then the \textit{sim}-AD chain models the qubit-wise decay of excitations to the environment register. This is set up such that the excitations in more significant phase qubits have a higher probability of decaying into their respective environment qubit. Since smaller eigenvalues are represented by larger phases, they are more likely to induce a decay, and are thus more severely damped when the measurement of the environment register does not find any excitation.

\subsubsection{Reverse operations and measurement}

The final step is the uncomputation of the phase register through the reversely applied QPE module, engendering the product state  $|\psi_4\rangle \approx |0\rangle^l|0\rangle^l|x(t)\rangle$. Moreover, in Fig. \ref{fig:Fig1}, the measurements corresponding to the $AD$ stage are pushed back to the end as it is customary in quantum circuits. This does not, however, alter our calculations, nor does it temper with the probability of measuring all $0$'s (i.e. the probability of success $p_s$).

The final link to the original problem is guaranteed by noting $ x(t) \dot{=} \sqrt{\gamma(t)}\:\big|\big|x_0\big|\big|e^{||A||t}\:|x(t)\rangle$. In order to calculate $\gamma(t)$ term-by-term by its expression given above, one should find the components $(c_i)_{1\leq i\leq N}$ and the eigenvalues $(a_i)_{1\leq i\leq N}$, which is just as hard as solving the original HLDE \cite{dubious}. However, this apparent obstacle is suppressed if one notices that $p_s = \gamma(t)$ and thus that $\gamma(t)$ can be estimated with the quantum algorithm itself and does not need to be classically pre or post calculated, much like the normalisation constant in \cite{harrow2009quantum}. 

\section{Setting the lower-bound on $p_S$}
\label{APC}

In cases where $p_S$ goes to zero as $t$ increases or in cases where the stationary value (lower-bound) of $p_S$ is deemed too low for some given application, one can opt to solve the following alternative HDLE problem:
\begin{equation}
    \label{eq:problem}
    \begin{cases}
    \frac{d}{dt}
    \begin{bmatrix}
    y(t) \\
    x(t)
    \end{bmatrix}

    = \begin{bmatrix}
    ||A||\:\mathbb{I}_N & 0 \\
    0  & A
    \end{bmatrix}_{2N\times 2N} \begin{bmatrix}
    y(t) \\
    x(t)
    \end{bmatrix} \\
        \begin{bmatrix}
    y(0) \\
    x(0)
    \end{bmatrix} =     \begin{bmatrix}
    y_0 \\
    x_0
    \end{bmatrix}_{2N\times 1} 
    \end{cases}
\end{equation}

This is equivalent to calculating simultaneously the evolution of two non-interacting systems, one that follows $\frac{dx(t)}{dt} = Ax(t)$ and one that follows $\frac{dy(t)}{dt} = ||A||y(t)$, the former being the one we are interested in and the latter only being used to increase the lower bound on $p_S$. 

The reasoning behind this method is to assure the existence of a (at least $N$-degenerate) subspace associated to the eigenvalue $||A||$, such that the weight decomposition of the normalised initial condition over this space is at least $\frac{||y_0||}{\sqrt{||x_0||^2+||y_0||^2}}$. This implicates the existence of a non-vanishing term - which also plays the role of a lower bound - in the expression of $\gamma(t)$.  

Indeed, if one takes $y_0 \in \{r\in \mathbb{C}^N\big| ||r||\geq ||x_0||\}$, then it is guaranteed that $p_S \geq 0.5$ for any $t\in \mathbb{R}_+$. In terms of implementation, this is payed off by adding an extra qubit to the work register, doubling the dimension of the associated Hilbert space. If one aims to extract a global characteristic of $x(t)$ through an observable, the choice of $y_0$ must be made accordingly; for instance, if the chosen global characteristic is the absolute average  $\big|\frac{1}{N}\sum_{i=1}^N(x(t))_i\big|$, then $y_0$ should be set such that its absolute average is $0$, thus not interfering in the final result.

Moreover, since this new larger matrix inherits the very same eigenvalue distribution of $A$, the other registers' sizes can be chosen as if the original problem was being tackled, hence yielding the same result in both cases if all measurement outcomes are $0$.

\section{Time complexity analysis}
\label{sec:complexity}

As our algorithm makes use of a QPE module for decomposing the initial condition state vector $|x_0\rangle$ into the eigen-basis of $A$, one can observe parallelism of the non-unitary operations realised for all resulting subspaces, which are each conditioned on the associated entangled state $|\tilde{\phi}_i\rangle$ in the phase register. Under such a scheme, one needs not to explicitly calculate the eigenvectors and eigenvalues of $A$, as would be required by a general classical approach. Hence, we deem it natural to encounter a quantum speed-up on the problem's dimension. This will be explicitly shown in this section through the inspection of all underlying modules. We also conclude on the limitations imposed on the complexity for the time variable $t$.

\subsection{Potential speed-up for scaling $N$}
There are two main ways to address the IC preparation stage: either it hints at another section of a larger routine (in which case the algorithm here described is simply a module) or an efficiently-implemented unitary gate $U_{x_0}$ should be applied in order to load $|x_0\rangle$ starting from some initial state $|\phi\rangle$, more closely in accordance with Fig. \ref{fig:Fig1}. In this second case, the loading of classical data can be thought of as some standard state-preparation method such as \cite{grover2002creating} or through  a quantum random access memory (qRAM) approach \cite{giovannetti2008architectures}. Indeed, when quantum memory is established, the complexity for loading the data into a qRAM can be about $\mathcal{O}(\log(N))$ \cite{giovannetti2008quantum}.

As for the QPE module, the implementation of $\mathcal{M}$ translates to the problem of Hamiltonian Simulation of $A$ for the simulated interval of time $\tau \equiv 2\pi/||A||$, which can be done efficiently if $A$ is efficiently row-computable and $s$-sparse (i.e. if each row has at most $s$ non-zero entries). More explicitly, the QPE requires the implementation of the family of unitary operators $(e^{-i2\pi A\tau 2^j})_{0\leq j \leq l-1}$, applied to the work register. Several approaches for sparse Hamiltonian Simulation were proposed in past years \cite{berry2007efficient, berry2009black, berry2015simulating, berry2015almost}. A known lower bound for the query complexity of this problem is $\Omega(s||A||_{\max}\tau + \log(1/\epsilon_H)/\log \log(1/\epsilon_H))$, and an optimal method in all parameters of interest has already been introduced \cite{low2017optimal}. The error introduced by this step ($\epsilon_H$) will not be considered, as the dominant source of error in the context of our calculations will be the non-exact phase estimation, hence the adoption of the soft complexity $\tilde{\mathcal{O}}(s||A||_{\max}\tau)$ for subsequent considerations.

In our algorithm, the total simulated time interval for implementing all the controlled-$\mathcal{M}$ gates within the QPE module amounts for $(L-1)\tau$ and thus $\mathcal{O}(L/||A||)$. The subsequent inverse Quantum Fourier Transform, only takes $\Theta(l^2)$ steps, which can be efficiently implemented \cite{kim2018efficient}. Similarly, the \textit{sim}-AD modules contribute with a total count of $\Theta(l)$ gates applied to $2$ qubits each. Thus, the total run-time complexity of the algorithm can be expressed as:
\begin{equation}
\label{Eq:Complex}
    \mathcal{O}\big(\log(N) + sL + \log^2(L) + \log(L)\big)
\end{equation}

We now derive, in a simple way, the complexity introduced by $L$, given the problem's parameters and some allowed additive error $\epsilon$ when approximating $x(t)$. This approach will only acknowledge the error introduced by the best possible approximation of each eigenvalue, given that a far more detailed approach (considering the totality of components in each $|\tilde{\phi}_i\rangle$) is given in Appendix \ref{tech_err}. This second approach yields the same result here depicted, albeit slightly more complete with the addition of the condition number of $A$ as a new term in the complexity expression.

Let $b_i \equiv \big( 1 - \frac{j}{L}\big)||A||$, for some $j \in \{0, 1...L-1\}$, be the best valid approximation of some eigenvalue $a_i$. Then, the algorithm errs by at most $\mathcal{O}(||A||/L)$ when estimating $a_i$ as $b_i$. This error is propagated to the exponential terms ($\epsilon_i \equiv |e^{b_it}-e^{a_it}|$) as follows:
\begin{equation*}
 \epsilon_i \leq e^{a_it}|e^{\frac{||A||t}{L}} - 1| \leq e^{||A||t}\big(e^{\frac{||A||t}{L}} - 1\big)
\end{equation*}

Solving for $L$ and using the asymptotic behaviour $1/\ln(1+ae^{-bx}) = \Theta(e^{bx}/a)$, found through the first order Taylor series approximation of $\ln(1+y)$ around $y=0$: 
\begin{equation*}
 L = \mathcal{O}\bigg(\frac{||A||te^{||A||t}}{\epsilon_i}\bigg)
\end{equation*}

More globally, if $\epsilon$ is the global additive error allowed to $x(t)$, it suffices to take $||\epsilon_i|| \equiv ||\epsilon||/||x_0||$, for all $i \in \{1, 2...N\}$. Thus:
\begin{equation}
\label{Eq:bigL}
 L = \mathcal{O}\bigg(\frac{||A||te^{||A||t}||x_0||}{\epsilon}\bigg)
\end{equation}

By combining Eq. \ref{Eq:Complex} and Eq. \ref{Eq:bigL}, if $s$, $||x_0||$ and $1/\epsilon$ are polylogarithmic in $N$ and if $||A||t = \mathcal{O}(1)$ \footnote{A less restrictive and still valid condition is $||A||t = \mathcal{O}(\log^{k_1} \log^{k_2} N)$ for some exponents $k_1$ and $k_2$.}, the algorithm reaches an exponential speedup compared to classical methods, in relation to the size of the problem $N$. 

\subsection{Limitations on the complexity for scaling $t$}

Whereas granting the aforementioned speedup in relation to the problem's size $N$, Eq. \ref{Eq:Complex} and Eq. \ref{Eq:bigL} imply an exponential complexity in $||A||t$. In this section, we present two complementary arguments for why this cannot be enhanced for any similar algorithm built from applying QPE with Hamiltonian Simulation.

Firstly, since the QPE module implicates an additive error when approximating the eigenvalues and given that this error scales linearly in relation to $1/L$, the Hilbert space dimension $L$ must scale exponentially with $t$ in order to keep the exponential terms $(e^{a_it})_{1\leq i\leq N}$ under some error boundary.

Secondly, It should be noted that the only term in Eq. \ref{Eq:Complex} that takes linear $L$ is the one related to the Hamiltonian Simulation, hence dominating the run-time when it comes to scaling $t$. This complexity, however, cannot be vanquished, coming as a direct result of the No-Fast-Forwarding Theorem \cite{berry2007efficient}, which yields that the simulation of a sparse Hamiltonian cannot be achieved in sub-linear complexity in relation to the simulated time, which in our construction is itself $\mathcal{O}(L)$ - when one accounts for the accumulated simulated time among all controlled-$\mathcal{M}$ gates in the QPE module.

\section{Conclusion}

In summary, the quantum circuitry presented here succeeds in solving Homogeneous Linear Differential Equations with a potentially exponential speed-up whilst compared to any classical method, concerning the problem's dimension $N$. Although no such speed-up is observed for other parameters (perhaps most importantly for $t$), we have provided the underlying results that yield the impossibility to perform better with any similar algorithm based on QPE with Hamiltonian Simulation. Also importantly, features such as the dominance of at most $2$-qubits elementary gates and the low connectivity requirements between the phase and environment registers boost the algorithm's appeal for scalable experimental implementations. 

The usage of a \textit{sim}-$AD$ chain also bears novelty by allowing the solution to be found through an open quantum system-inspired scheme, creating a subspace where each state of the computational basis is adequately decayed thanks to the tuning of a set of parametric gates. We stress that such a strategy does not demand real open system dynamics. Despite its non-interferometric nature, we have shown that one can also foster the decaying property of Amplitude Damping to favour the correct output (Section \ref{APC}), which ultimately imposes a $t$-independent minimal probability of success for every single realisation of the algorithm, independently of the underlying HDLE problem. Such a scheme may create a new perspective for designing quantum algorithms based on open-systems-inspired operations, particularly for applications where specific non-unitary transformations are needed.   

\section{Acknowledgements}

F.B. is supported by Instituto Nacional de Ci\^encia e Tecnologia de Informa\c c\~ao Qu\^antica (CNPq INCT-IQ 465469/2014-0), Brazil. The authors would like to thank Professor Diogo Soares-Pinto for the fruitful discussions.

\appendix

\section{A closer look into the non-unitary section}
\label{closerlook}

The chore of the algorithm here presented lies in the \textit{sim}-AD stage, as it enables the non-unitary calculation using exclusively two common types of $1$-qubit controlled gates, and with a gate depth that only scales linearly with the register size $l$ (hence logarithmically in the allowed error $1/\epsilon$). In this appendix, we explicitly derive Eq. \ref{eq:expAPA}, while also taking the opportunity to justify our choice of operator $\mathcal{M}$ for the previous QPE step. 

As presented before, the correct transformation over the phase register is applied if and only if the measurement outcomes are all $0$. In this case, we know that the output state of the QPE stage $|\psi_2\rangle$ has been taken into the state $\frac{G_0|\psi_2\rangle}{\langle \psi_2 |G_0^\dagger{G_0| \psi_2 \rangle}}$.  An uncomplicated way to find the expression for $G_0$ which was used in this article is to look into how it operates over a state of the computational basis $|j\rangle$. First of all, it should be noted that the $E_0^k$ operators can be rewritten as $E_0^k = \sum_{n=0}^1e^{-||A||2^knt/L}|n\rangle\langle n|$. If $j$ is expressed in its binary form $(j_{l-1}j_{l-2}...j_0)_2$, it becomes clear that:
\begin{eqnarray*}
    G_0|j\rangle &=& \bigotimes_{k=0}^{l-1}\sum_{n=0}^1e^{-||A||2^knt/L}|n\rangle\langle n|j_k\rangle \\
   &=& \bigotimes_{k=0}^{l-1}e^{-||A||2^kj_kt/L}|j_k\rangle\\
  &=& e^{-||A||\big(\sum_{k=0}^{l-1}j_k2^k\big)t/L}|j\rangle\\
 &=& e^{-||A||jt/L}|j\rangle
\end{eqnarray*}

Thus, $G_0$ can be generally expressed as the following sum of projectors in the computational basis:
\begin{equation}
\label{eq:apG0}
    G_0 =\sum_{j=0}^{L-1} e^{-||A||jt/L}|j\rangle\langle j|
\end{equation}

Recalling that the states $|j\rangle_{0\leq j\leq L-1}$ code the phases of the spectrum of $\mathcal{M}$, Eq. \ref{eq:apG0} shows that the damping effect increases (i.e. the exponent's absolute value increases linearly) as we scan the complex unit circle in an anti-clockwise manner starting from the rightmost extremity, (recall Fig. \ref{fig:Fig3_ant}). While $|0\rangle$ suffers no damping, the largest allowed damping is $e^{-||A||(L-1)t/L} \xrightarrow{L\rightarrow \infty} e^{-||A||t}$. This should be enough to justify our choice of unitary operator $\mathcal{M}$ as it linearly wraps the eigenvalues of $A$ around the complex unit circle in such a way that subspaces related to the smallest eigenvalues suffer more damping - i.e. are associated to larger phases within the $[0, 2\pi[$ range.


The effect of $G_0$ on $|\psi_2\rangle$ from Eq. \ref{eq:QPEform2} can now be calculated:
\begin{eqnarray*}
G_0|\psi_2\rangle &=&  \sum_{i=1}^{N}c_i\sum_{j=0}^{L-1} e^{-||A||jt/L}|j\rangle\langle j| d_i \rangle|v_i\rangle\\
&=& \sum_{i=1}^{N}c_i e^{-||A||\big(\lfloor L(1-a_i/||A||)\rceil \pmod L\big)t/L}\big| d_i   \rangle|v_i\rangle
\end{eqnarray*}

We can dispose of the modular notation by assuming that there is no $i\in \{1,2...N-1\}$ such that  $\lfloor L(1-a_i/||A||)\rceil = L$; note that if the negation of this statement is true, then we could observe an additive error which grows as fast as $O(e^{||A||t})$ - since at least one subspace associated to some $a_i < ||A||$ would suffer no damping. This is particularly bound to happen if $L$ is not sufficiently large for a particular high value of the condition number ($\kappa$) of $A$. This is the case because $m_1$ approximates $0$ in Fig. \ref{fig:Fig3_ant} through the fourth quadrant as $\kappa$ increases. This consideration will be further developed in Appendix \ref{tech_err}, where it will ultimately lead to the introduction of $1/\kappa$ in a revised form of Eq. \ref{Eq:bigL}.

Without the modular notation, we will now provide a guideline for proving that $G_0|\psi_2\rangle \xrightarrow[]{L \rightarrow \infty} \sum_{i=1}^{N}c_i e^{-(a_N-a_i)t}\big| d_i   \rangle|v_i\rangle$, and thus that the precision can always be enhanced by augmenting the size of the phase register. Indeed, for any $z\in [0, 1[$:

\begin{equation*}
    \bigg|\frac{\lfloor Lz\rceil}{L} - z\bigg| \leq \bigg|\frac{Lz + 1}{L} - z\bigg| = 1/L
\end{equation*}

Which implies $\lfloor L(1-a_i/||A||)\rceil/L\xrightarrow[]{L \rightarrow \infty} 1-a_i/||A||$. The convergence that we aim to prove then follows from the continuity of the associated exponential functions. 

\section{Generalised algorithm for any Hermitian matrix}
\label{APB}

\begin{figure}[H]
    \centering
    \includegraphics[scale = 0.3]{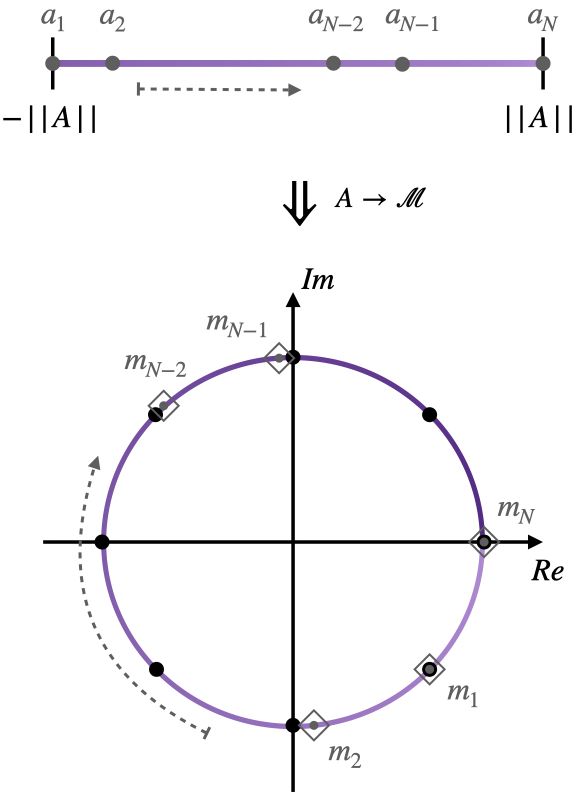}
    \caption{Illustrative example of the eigenvalues mapping between $A$ and $\mathcal{M}$, for the less restrictive Hermitian $A$ formulation. The $[-||A||, ||A||]$ interval is mapped to the complex phase interval $\big[0, 2\pi(L-1)/L\big]$. The black filled dots around the circle represent the allowed values of damping, in this case for $l=3$.}
    \label{fig:Fig4}
\end{figure}

As suggested in this article, our algorithm can be extended to solve HDLE problems with any Hermitian matrix $A$. In fact, this task can be achieved by simply modifying a few parameters of the equivalent quantum circuit - as the number and size of quantum registers and applied quantum gates remain untouched. 

The first modification comes to the QPE stage: while the $\bar{A} \equiv A/||A||$ operator is still defined the same way, the new form of the unitary operator $\mathcal{M}$ should be constructed such that the interval $[-||A||, ||A||]$ is mapped in one single lap around the unit circle, without overlapping both ends. The revised operator is given by:
\begin{equation}
   \mathcal{M} \equiv e^{-i2\pi\big(\frac{L-1}{2L}\bar{A} +\frac{L+1}{2L}\mathbb{I}_N\big)} 
\end{equation}

One may refer to Fig. \ref{fig:Fig4} for a visual interpretation of $\mathcal{M}$. A brief side-note should be made about the implementation for the equivalent circuit: $\mathcal{M}$ can be achieved by Hamiltonian simulation of $\frac{L-1}{2L}\bar{A} +\frac{L+1}{2L}\mathbb{I}_N$, which, for a $s$-sparse matrix $A$, is at most  $(s+1)$-sparse. As a direct result, this stage keeps the same overall complexity as presented in Section \ref{sec:complexity} for the more restrict formulation.

The output of the QPE stage will still follow the same form depicted in Eq. \ref{eq:QPEform1}, where if $|\tilde{\phi}_i\rangle = \sum_{j=0}^{L-1}\beta_{i;j}|j\rangle$, then $|\beta_{i;j}|$ takes a high value if and only if $j \approx \frac{L-1}{2}(1-a_i/||A||) \pmod L$. The mod notation is immediately dropped with no loss of generality since $\min_{1\leq i\leq N} (a_i) \geq -||A||$. Thus, following the same assumptions described in Section \ref{sec:QPE}:
\begin{equation}
\label{eq:QPEformgeneralized}
|\psi_2\rangle \approx
\sum_{i=1}^{N}c_i\bigg|  \big\lfloor\frac{L-1}{2}(1-a_i/||A||)\big\rceil \bigg\rangle |v_i\rangle
\end{equation}

The second and last modification to our algorithm comes to the rotation angles within the \textit{sim}-AD modules:
\begin{equation}
    \label{eq:thetasGen}
    \begin{cases}
    \forall k\in \{0,1...l-1\} \: \theta_k \equiv \arccos(2e^{-||A||2^{k+2}t/(L-1)}-1)
    \end{cases}
\end{equation}

By comparing Eq. \ref{eq:thetasGen} to Eq. \ref{eq:thetas}, the new expression for $G_0|\psi_2\rangle$ is:
\begin{equation*}
G_0|\psi_2\rangle =  \sum_{i=1}^{N}c_i\bigg(\sum_{j=0}^{L-1} e^{-2||A||jt/(L-1)}|j\rangle\langle j|\bigg) |d_i \rangle|v_i\rangle
\end{equation*}

If all measurements to the environment register result in $0$, the global output is once again achieved as $|\psi_3\rangle \approx \frac{1}{\sqrt{\gamma(t)}}  
\sum_{i=1}^{N}c_ie^{-(||A||-a_i)t}  | d_i \rangle|v_i\rangle$, with $\gamma(t) \equiv \sum_{i=1}^{N}|c_i|^2e^{-2(||A||-a_i)t}$. However, a new subtlety compared to the more restrictive case is that $a_N$ can now be strictly smaller than $||A||$. Under such circumstances, $p_S = \gamma(t) \xrightarrow[]{t \rightarrow\infty} 0$, i.e. the probability of success necessarily vanishes as $t$ increases, regardless of the initial conditions. Nevertheless, this can be simply resolved by the approach presented on Section \ref{APC}.

\section{In-depth analysis of the phase register size}
\label{tech_err}

In Section \ref{sec:complexity}, a simple calculation of the complexity of $L$ was given (Eq. \ref{Eq:bigL}). In this appendix, we present a more thorough analysis of the phase register size, which will ultimately lead to the same conclusion on the complexity, although with a new parameter - the condition number $\kappa$. We start by looking into the particular case of one single eigenvalue $a_i$, i.e. $|x_0\rangle \equiv |v_i\rangle$,  and easily extend it to any initial condition. Indeed, one can revisit Eq. \ref{eq:QPEform1} and the exact expression for $|\tilde{\phi}_i\rangle$ given in Section \ref{sec:QPE}. The latter should be modified upon relabeling the phase register as shown below \cite{nielsen2002quantum}:
\begin{equation*}
\begin{cases}
    |\tilde{\phi}_i\rangle = \sum_{k=-L/2+1}^{L/2}\alpha_{j|i}|j\rangle \\
    \\
    \text{with } \alpha_{j|i} = \frac{1}{L}\sum_{k=-L/2+1}^{L/2}\bigg[e^{i2\pi\big(\phi_i - \frac{d_i+ j}{L}\big)}\bigg]^k

\end{cases}
\end{equation*}

Where $\phi_i \equiv (1-\bar{a}_i)$, and $d_i/L \in [0, 1[$ is the best $l$ bits approximation of $\phi_i \pmod 1$. Continuing with the procedure, the \textit{sim}-AD modules and the inverse QPE, upon a successful final measurement and after the adequate rescaling, will result in:
\begin{equation*}
\begin{cases}

    \tilde{x}(t) =  \bigg|\sum_{k=-L/2+1}^{L/2}\alpha_{j|i}e^{b_{j|i}t}\bigg|v_i \\
    \\
     \text{with } b_{j|i} \equiv ||A||\big(1-\frac{d_i+j}{L}\big)

\end{cases}
\end{equation*}

Where the absolute value is introduced in order to discard a global phase that may appear. On the other hand, the analytical solution simply yields $x(t) = e^{a_it}v_i$. We also introduce the overall additive error $\epsilon_i = \tilde{x}(t) - x(t)$ of the algorithm. 

In order to proceed, the $j$ sub-indexes shall be clustered into $2$ groups, by verifying if they respect the following criteria: $\big|\phi_i - \frac{d_i+j}{L}\big| < P$, for some precision $P\in ]0, 1[$ which will be later specified. Since $d_i/L \approx \phi_i$, this criteria can be simplified as $|j|\leq p$, with $p\equiv \lceil LP \rceil$, which stipulates the following upper bound on $e_i$:
\begin{equation*}
\begin{cases}

    ||\epsilon_i||^2 \leq ||\epsilon_{QPE}||^2 + ||\epsilon_{Pr}||^2 \\
    \text{with }||\epsilon_{QPE}||^2 \equiv \bigg|\bigg|\sum_{j\not\in \{-p,...p-1, p\}}|\alpha_{j|i}|e^{b_{j|i}t}v_i\bigg|\bigg|^2 \\ 
    \text{and }||\epsilon_{Pr}||^2 \equiv\bigg|\bigg|\bigg(\sum_{j=-p}^{p}|\alpha_{j|i}|e^{b_{j|i}t} - e^{a_it}\bigg)v_i\bigg|\bigg|^2
\end{cases}
\end{equation*}

We will hereby refer to the first term as the QPE error $||\epsilon_{QPE}||^2$ and the second term as the precision error $||\epsilon_{Pr}||^2$.

\subsection{QPE error}

The QPE error can be bounded as follows:
\begin{equation*}
    ||\epsilon_{QPE}||^2 \leq \bigg(\sum_{\substack{-L/2 < j < -p \\ p <j\leq L/2}}|\alpha_{j|i}|^2\bigg)e^{2||A||t}
\end{equation*}

This sum of squared probability has $\frac{1}{2(p-1)}$ as a valid upper bound \cite{nielsen2002quantum}. Using $p\equiv \lceil 2^lP \rceil$:
\begin{equation*}
    ||\epsilon_{QPE}||^2 \leq \frac{e^{2||A||t}}{2(2^lP-1)}
\end{equation*}

Solving for $l$, one may conclude that it suffices to use a phase register as large as:
\begin{equation*}
    l \leq \log_2\bigg(\frac{e^{2||A||t}}{2||\epsilon_{QPE}||^2} + 1 \bigg) + \log_2\bigg(\frac{1}{P}\bigg)
\end{equation*}
And, thus:
\begin{equation}
\label{bigO_QPE}
    l = \mathcal{O}\bigg(||A||t + \log\bigg(\frac{1}{||\epsilon_{QPE}||}\bigg) + \log\bigg(\frac{1}{P}\bigg)\bigg)
\end{equation}

Eq. \ref{bigO_QPE} will be revisited when we set a bound on $P$, which will come naturally as we delve into the precision error analysis.

\subsection{precision error}
For every value of $j$ within $\{-p, -p+1...p-1, p\}$, the individual error introduced  can be bounded as follows:
\begin{equation*}
    |e^{b_{j|i}t} - e^{a_it}|^2 \leq |e^{b_{-p|i}t} - e^{a_it}|^2
\end{equation*}

The right-hand side can then be rewritten as $|e^{b_{-p|i}t} - e^{b_{0|i}t}|^2 + |e^{b_{0|i}t} - e^{a_it}|^2$ through triangular inequality. The first term in this sum translates to how close the most extreme accepted approximation ($e^{b_{-p|i}t}$) of the exact exponential term is from the best possible approximation ($e^{b_{0|i}}$) of the exponential term, and the second term corresponds to how close this best approximation meets the exact solution. While the former quadratic error will lead to a bound on $P$, the latter will result in another bound on the phase register size $l$.

Indeed, the first aforementioned quadratic error can be rewritten as follows:
\begin{equation*}
 ||\epsilon_{Pr;I}||^2 \leq e^{2b_{0|i}t}|e^{p/L} - 1|^2 \leq e^{2||A||t}|e^{\lceil LP \rceil/L} - 1|^2
\end{equation*}

Which, simplifying and solving for $P$, yields:
\begin{equation*}
 \frac{1}{P} \leq \frac{1}{\ln(1+||e_{Pr;I}||e^{-||A||t})}
\end{equation*}

And, once again using $1/\ln(1+ae^{-bx}) = \Theta(e^{bx}/a)$: 
\begin{equation}
\label{Eq:bigOdeP}
     \frac{1}{P} = \mathcal{O}\bigg(\frac{e^{||A||t}}{||\epsilon_{Pr;I}||}\bigg)
\end{equation}

One can also advocate for the inclusion of $\kappa$, the condition number of $A$, in the expression of $1/P$. Fig. \ref{fig:Fig3_ant} gives some visual input to why this is the case. Indeed, we should not allow $P$ to be bigger than $a_1/a_N \equiv 1/\kappa$ as this would imply that the region covered by the $\{-p, -p+1...p-1, p\}$ indexes (normally associated to significant amplitudes $|\alpha_{j|1}|$ according to our construction) would cross over to the first quadrant, where terms suffer much less damping, generating an error that could grow as fast as $\mathcal{O}(e^{||A||t})$. Hence, it would also be more suitable to set $1/P = \mathcal{O}(\kappa e^{||A||t}/||\epsilon_{Pr;I}||)$.

Moving to the second quadratic error term, it should be noted that the difference between the exact eigenvalue $a_i$ and its best allowed approximation $b_{0|i}$ cannot be larger than $||A||/2L$. With that being said:
\begin{equation*}
 ||\epsilon_{Pr;II}||^2 \leq e^{2a_it}|e^{\frac{||A||t}{2L}} - 1|^2 \leq e^{2||A||t}\big|e^{\frac{||A||t}{2L}} - 1\big|^2
\end{equation*}

And upon some very similar calculations to those that preceded Eq. \ref{Eq:bigOdeP}:
\begin{equation}
\label{Eq:bigO_Pr}
 l = \mathcal{O}\bigg(||A||t + \log\bigg(\frac{1}{||\epsilon_{Pr;II}||}\bigg) \bigg)
\end{equation}

Where a slower growing logarithmic term on $||A||t$ was suppressed. 

\subsection{Concluding on the Phase Register Size}

Putting Eq. \ref{bigO_QPE}, \ref{Eq:bigOdeP} and \ref{Eq:bigO_Pr} together, we stipulate that the overall expression for the phase register size so far is $l = \mathcal{O}\bigg(||A||t +\log\bigg(\frac{1}{||\epsilon_i||}\bigg) + \log(\kappa)\bigg)$.

As we finally lift the constraint of looking into one single eigenvalue and move to any initial condition $x_0 = \sum_{i=1}^N\tilde{c}_i v_i$, it suffices to verify that the total error is bounded by $||\epsilon||^2\leq  ||x_0||^2\max_{1\leq i\leq N}(||\epsilon_i||^2) $, and thus that one can choose $||e_i|| \equiv ||\epsilon||/||x_0||$ in our last expression in order to conclude about the asymptotic behaviour imposed over $l$:
\begin{equation}
\label{Eq:final}
     l = \mathcal{O}\bigg(||A||t + \log(||x_0||) + \log\bigg(\frac{1}{||\epsilon||}\bigg) + \log(\kappa)\bigg)
\end{equation}

%

\end{document}